\documentclass[sigconf]{acmart}

\usepackage{booktabs} 
\usepackage{url}
\usepackage{hyperref}
\usepackage{tabularx}
\usepackage[linesnumbered, ruled, vlined]{algorithm2e}
\usepackage{float}
\usepackage{setspace}
\usepackage{multirow}

\let\oldnl\nl
\newcommand{\nonl}{\renewcommand{\nl}{\let\nl\oldnl}}

\usepackage{textcomp} 
\newcommand{\qdist}[1]{\ifmmode\langle#1\rangle\else\textlangle#1\textrangle\fi}

\newlength\mylen

\renewcommand\footnotetextcopyrightpermission[1]{} 

\acmConference[I3D '19]{ACM SIGGRAPH Symposium on Interactive 3D Graphics and Games}{21-23 May 2019}{Montreal, Quebec, Canada}

\begin{document}
\title{Computing Three-dimensional Constrained Delaunay Refinement Using the GPU}

\author{Zhenghai Chen}
\affiliation{%
  \department{School of Computing}
  \institution{National University of Singapore}
}
\email{chenzh@comp.nus.edu.sg}

\author{Tiow-Seng Tan}
\affiliation{%
  \department{School of Computing}
  \institution{National University of Singapore}
}
\email{tants@comp.nus.edu.sg}


\begin{abstract}
We propose the first GPU algorithm for the 3D triangulation refinement problem. For an input of a piecewise linear complex $\mathcal{G}$ and a constant $B$, it produces, by adding Steiner points, a constrained Delaunay triangulation conforming to $\mathcal{G}$ and containing tetrahedra mostly 
of radius-edge ratios smaller than $B$.
Our implementation of the algorithm shows that it can be an order of magnitude faster than the best CPU algorithm while using a similar amount of Steiner points to produce triangulations of comparable quality. 
\end{abstract}

%
%



\maketitle

\section{Introduction}

Constrained Delaunay triangulations (CDTs) are used in various engineering and scientific applications, such as finite element methods, interpolation etc. Such a CDT, in general, is obtained from a so-called \emph{piecewise linear complex} (PLC) $\mathcal{G}$ containing a point set $P$, an edge set $E$ (where each edge with endpoints in $P$), and a polygon set $F$ (where each polygon with boundary edges in $E$).
All vertices, edges and polygons of $\mathcal{G}$ also appear in $\mathcal{T}$ as vertices, union of edges, and union of triangles, respectively; we also say $\mathcal{T}$ \emph{conforms} to $\mathcal{G}$ in this case.
For our discussion, we call an edge in $E$ a \emph{segment}, an edge in $\mathcal{T}$ which is also a part (or whole) of some segment a \emph{subsegment}, and 
a triangle in $\mathcal{T}$ which is also a part (or whole) of some polygon of $F$ a \emph{subface}.

For a given constant $B$ and a CDT $\mathcal{T}$ of $\mathcal{G}$ as input, the constrained Delaunay refinement problem is to add vertices, called \emph{Steiner points}, into $\mathcal{T}$ to eliminate or split most, if not all, bad tetrahedra to generate a new CDT of $\mathcal{G}$. (A tetrahedron $t$ is \emph{bad} if the ratio of the radius of its circumsphere to its shortest edge is larger than $B$.) A solution to the problem should also aims to add few Steiner points. 
The {\tt TetGen} software by \citet{Si15} is the best CPU solution known to the problem. 
It, however, can take a significant amount of time of minutes to hours to compute CDTs for some typical inputs from applications. 
We thus explore the use of GPU to address this problem.


\section{Our Proposed Algorithm}

Our proposed algorithm {\tt gQM3D} follows the general Delaunay refinement paradigm where subsegments, subfaces and bad tetrahedra, collectively called \emph{elements}, are split in this order in many rounds until there are no more bad tetrahedra. Each round, the splitting is done to many elements in parallel with many GPU threads. The algorithm first calculates the so-called \emph{splitting points} that can split elements into smaller ones, then decides on a subset of them to be Steiner points for actual insertions into the triangulation $\mathcal{T}$. Note first that a splitting point is calculated by a GPU thread as the midpoint of a subsegment, the circumcenter of the circumcircle of the subface, and the circumcenter of the circumsphere of the tetrahedron. Note second that not all splitting points calculated can be inserted as Steiner points in a same round as they together can potentially create undesirable short edges in $\mathcal{T}$ to cause non-termination of the algorithm. So, the algorithm must filter away some splitting points.  

For a splitting point $p$, its \emph{Delaunay region} is the set of elements (subfaces or tetrahedra) who will become non-Delaunay (with their circumcircles or circumspheres, respectively, enclosing $p$) if $p$ is inserted as a Steiner point into $\mathcal{T}$. 
We know for two splitting points with disjoint Delaunay regions, their  insertions into $\mathcal{T}$ will not result in them forming an edge in $\mathcal{T}$ (while $\mathcal{T}$ is maintained as a constrained Delaunay triangulation at the end of each round). As such, and to achieve good speed up with using the GPU, our algorithm seeks to identify a large set of splitting points with mutually disjoint Delaunay regions in each round. So, the problem becomes how to identify disjoint Delaunay regions efficiently. 

The trivial way of one thread taking care of one splitting point to calculate its Delaunay region is inefficient as different threads can need vastly different amounts of computation to process Delaunay regions of different sizes. Instead, a good approach should deploy a number of threads in proportion to the size of a Delaunay region so each thread does more or less similar amount of work. Such a desirable regularized work approach is developed in our \emph{grow-and-blast} scheme as outline in the next paragraph.  

Initially, a thread is assigned to an element where the splitting point is located. This element is also a part of the Delaunay region of the splitting point. The thread then checks the neighbors (subfaces and tetrahedra) of this element to decide whether they are also a part of the Delaunay region of the splitting point. For such a neighbor, it is marked (grown) as a part of the Delaunay region, and a thread will be assigned to it to perform the similar kind of checking and marking subsequently. Having said this, when an element appears as a neighbor to many and is to be marked into more than one Delaunay regions, only one is allowed while others with predetermined lower priorities must be stop (blasted) and their corresponding splitting points filtered away. Those Delaunay regions remain are mutually disjoint, and their corresponding splitting points are inserted concurrently into $\mathcal{T}$ as Steiner points.

\begin{table*}[h]
  \centering
  \scalebox{0.70}
  {
    \begin{tabular}{| c | c | c | c | c | c | c | c | c | c | c | c | c | c | c | c |}
      \hline
          $\gamma$ & \multicolumn{3}{c|}{0.05} & \multicolumn{3}{c|}{0.10}
          & \multicolumn{3}{c|}{0.15} & \multicolumn{3}{c|}{0.20} & \multicolumn{3}{c|}{0.25}
          \\
      \hline
      \multicolumn{1}{|r|}{algorithm}
      & \multirow{2}{*}{\emph{TetGen}}  & \multirow{2}{*}{\emph{gQM3D}}  & \multirow{2}{*}{\emph{gQM3D\textsuperscript{+}}}
      & \multirow{2}{*}{\emph{TetGen}}  & \multirow{2}{*}{\emph{gQM3D}}  & \multirow{2}{*}{\emph{gQM3D\textsuperscript{+}}} 
      & \multirow{2}{*}{\emph{TetGen}}  & \multirow{2}{*}{\emph{gQM3D}}  & \multirow{2}{*}{\emph{gQM3D\textsuperscript{+}}} 
      & \multirow{2}{*}{\emph{TetGen}}  & \multirow{2}{*}{\emph{gQM3D}}  & \multirow{2}{*}{\emph{gQM3D\textsuperscript{+}}}
      & \multirow{2}{*}{\emph{TetGen}}  & \multirow{2}{*}{\emph{gQM3D}}  & \multirow{2}{*}{\emph{gQM3D\textsuperscript{+}}}
      \\
      \multicolumn{1}{|l|}{$B$} & & & & & & & & & & & & & & &\\
      \hline
      \multicolumn{1}{|l|}{$1.4$} & & & & & & & & & & & & & & &\\
      Time (min) & 2.5 & 1.3 & 0.9 & 6.6 & 2.2 & 1.5 & 20.4 & 3.1 & 2.3
      & 28.6 & 3.9 & 2.9 & 53.4 & 4.5 & 4.0
      \\
      Points (M) & 0.95 & 0.93 & 0.93 & 1.52 & 1.49 & 1.50 & 2.63 & 2.59 & 2.61 &
      3.11 & 3.06 & 3.08 & 4.24 & 4.18 & 4.21
      \\
      Tets (M) & 5.98 & 5.85 & 5.88 & 9.58 & 9.40 & 9.44 & 16.68 & 16.37 & 16.45
      & 19.67 & 19.35 & 19.46 & 26.89 & 26.44 & 26.64
      \\
      Bad Tets & 401 & 308 & 376 & 1461 & 1416 & 1564 & 2160 & 2059 & 2156
      & 2885 & 2939 & 2894 & 3677 & 3395 & 3765
      \\
      \hline
      \multicolumn{1}{|l|}{$1.6$} & & & & & & & & & & & & & & &\\
      Time (min) & 1.6 & 1.3 & 0.7 & 4.1 & 2.2 & 1.3 & 12.8 & 3.1 & 2.2
      & 18.3 & 3.9 & 2.6 & 34.3 & 4.5 & 3.3
      \\
      Points (M) & 0.68 & 0.69 & 0.69 & 1.12 & 1.13 & 1.14 & 2.03 & 2.06 & 2.07
      & 2.39 & 2.44 & 2.45 & 3.33 & 3.39 & 3.41
      \\
      Tets (M) & 4.27 & 4.33 & 4.34 & 7.00 & 7.10 & 7.11 & 12.73 & 12.91 & 12.97 &
      15.06 & 15.29 & 15.36 & 20.94 & 21.28 & 21.40
      \\
      Bad Tets & 303 & 252 & 285 & 1279 & 1152 & 1245 & 1877 & 1725 & 1848
      & 2520 & 2355 & 2480 & 3235 & 2924 & 3264
      \\
      \hline
      \multicolumn{1}{|l|}{$1.8$}& & & & & & & & & & & & & & &\\
      Time (min) & 1.11 & 1.08 & 0.70 & 2.90 & 1.67 & 1.19 & 9.02 & 2.48 & 1.91
      & 12.76 & 3.29 & 2.71 & 24.13 & 4.63 & 3.04
      \\
      Points (M) & 0.56 & 0.57 & 0.58 & 0.92 & 0.95 & 0.95 & 1.73 & 1.79 & 1.79
      & 2.05 & 2.12 & 2.12 & 2.88 & 2.97 & 2.99
      \\
      Tets (M) & 3.46 & 3.57 & 3.58 & 5.74 & 5.93 & 5.93 & 10.76 & 11.10 & 11.14
      & 12.75 & 13.16 & 13.22 & 17.94 & 18.51 & 18.60
      \\
      Bad Tets & 251 & 229 & 252 & 1083 & 1467 & 1107 & 1599 & 1473 & 1582
      & 1998 & 2004 & 2025 & 2696 & 2484 & 2768
      \\
      \hline
      \multicolumn{1}{|l|}{$2.0$} & & & & & & & & & & & & & & &\\
      Time (min) & 0.84 & 1.00 & 0.58 & 2.21 & 1.81 & 1.04 & 6.86 & 2.62 & 1.77
      & 9.72 & 3.10 & 2.06 & 18.52 & 3.59 & 3.02
      \\
      Points (M) & 0.49 & 0.51 & 0.51 & 0.82 & 0.85 & 0.85 & 1.57 & 1.62 & 1.63
      & 1.86 & 1.92 & 1.93 & 2.63 & 2.73 & 2.74
      \\
      Tets (M) & 3.02 & 3.14 & 3.14 & 5.06 & 5.26 & 5.27 & 9.66 & 10.03 & 10.06
      & 11.48 & 11.89 & 11.94 & 16.25 & 16.85 & 16.91
      \\
      Bad Tets & 232 & 201 & 235 & 967 & 935 & 996 & 1381 & 1294 & 1397
      & 1746 & 1670 & 1759 & 2330 & 2149 & 2332
      \\
      \hline
    \end{tabular} 
  }
  
  \vspace{0.1in}
  \caption{Comparison among algorithms with 25K input points of the ball distribution. "Tets" denotes tetrahedra.}
  \label{3algo}
\end{table*}

\section{Experimental Results}
All experiments are conducted on a PC with an Intel i7-7700k 4.2GHz CPU, 32GB of DDR4 RAM and a GTX1080 Ti graphics card with 11GB of video memory. {\tt TetGen} is the main CPU software we use to compare with our {\tt gQM3D} implemented with CUDA programming model. 
During our experimentation, we notice {\tt gQM3D} does not have particular advantage over CPU approach for the initial part of the computation. We thus replace this part of {\tt gQM3D} by using {\tt TetGen} in CPU to obtain a variant called {\tt gQM3D}$^+$. 
We note that {\tt CGAL} \cite{cgal3d} and {\tt TetWild} \cite{Hu18} are not part of the comparison for now as they address a slightly different problem that allows output not conforming to the input PLCs. 
%
%

\begin{figure}[h]
	\centering 
	\includegraphics[scale = 0.41]{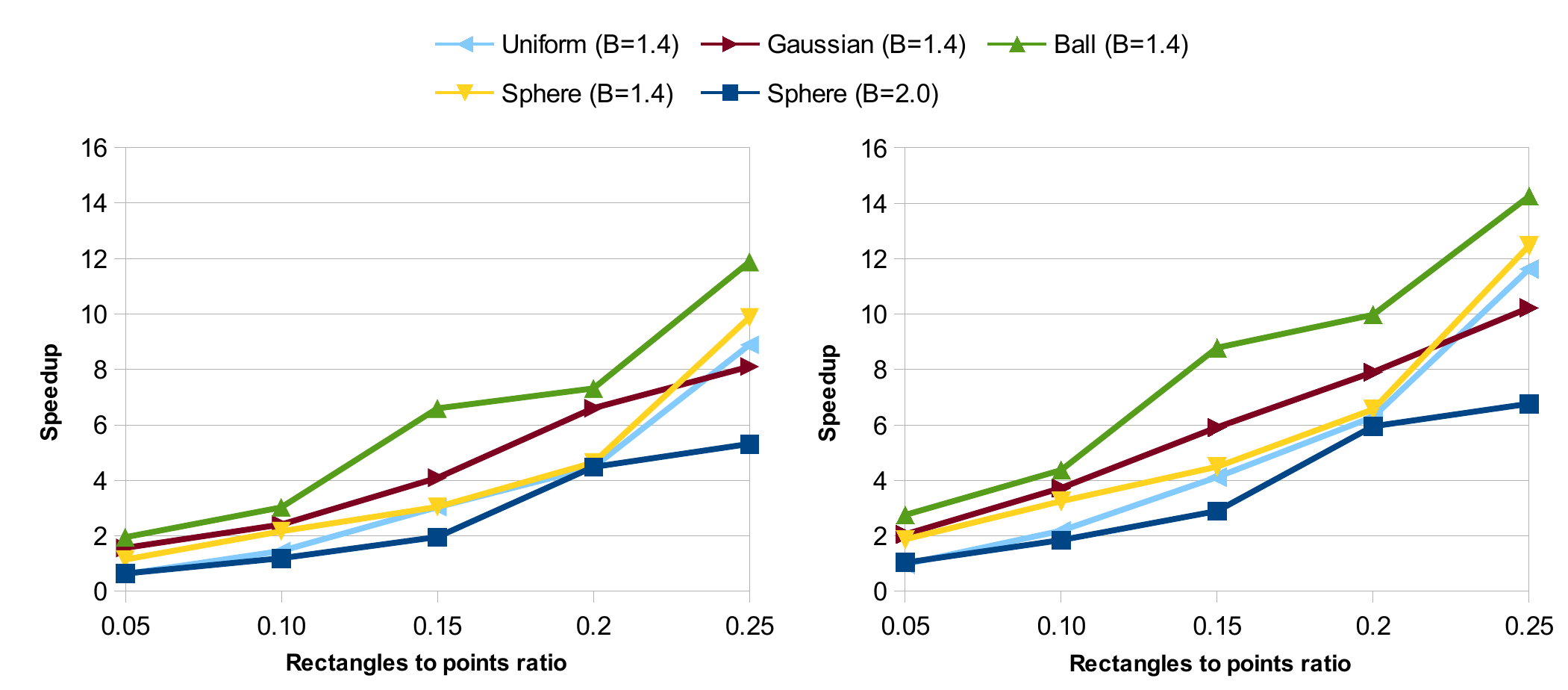}
	\caption{Speedup of {\tt gQM3D} (left) and {\tt gQM3D}$^+$ (right).}
	\label{fig:speedup}
\end{figure}

\begin{figure}[h]
	\centering 
	\includegraphics[scale = 0.4]{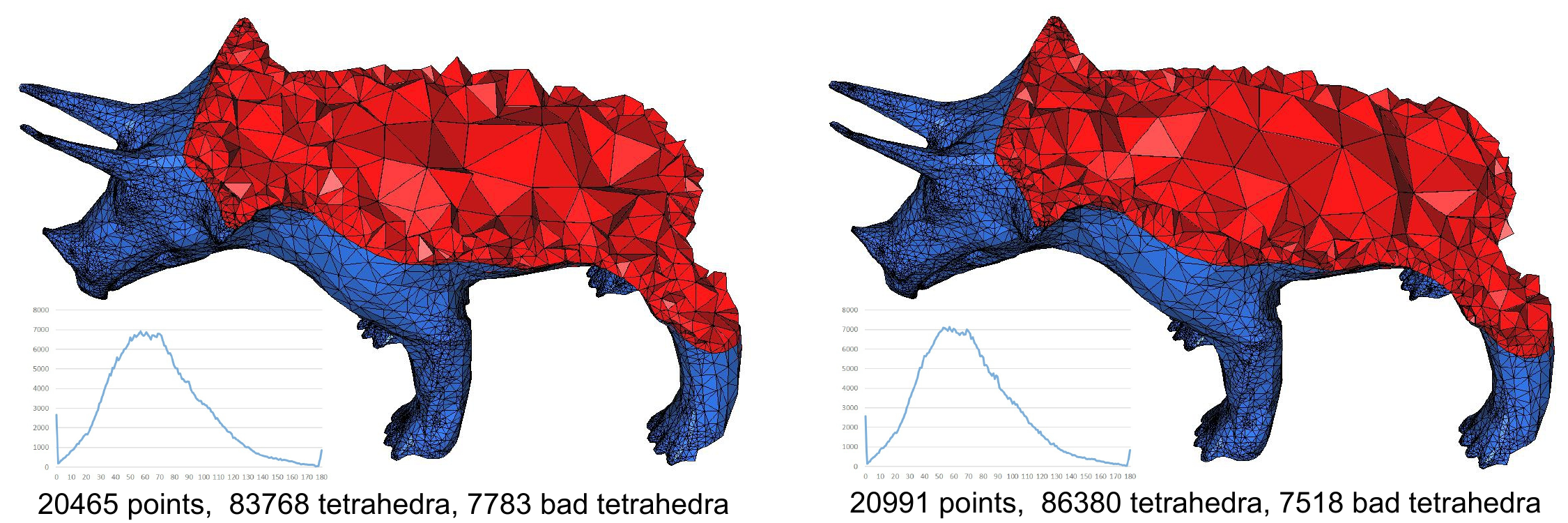}
	\caption{The output triangulations of a triceratops generated by {\tt TetGen} (left) and {\tt gQM3D} (right). }
	\label{fig:quality}
\end{figure} 

Table \ref{3algo} and Figure \ref{fig:speedup} report the running time and triangulation quality obtained with synthetic PLCs with points of different distributions. 
$\gamma$ is the ratio of the number of polygons (which are mainly rectangles) to the number of points in the input PLC.
Both {\tt gQM3D} and {\tt gQM3D}$^+$ can achieve speedup of an order of magnitude while generate outputs with similar sizes compared to that of {\tt TetGen}. Figure \ref{fig:quality} shows (cut-off views) the comparison of output triangulations of a real-world object for {\tt TetGen} and {\tt gQM3D}. The outputs have similar sizes with the latter having slightly more Steiner points but fewer bad tetrahedra. Both triangulations have similar distribution of dihedral angles (ranging from $0^\circ$ to $180^\circ$) as shown in the inserted line graphs and thus of equally good triangulations.

\section{Concluding Remarks}
We propose the first GPU algorithm for the constrained Delaunay refinement problem. It is designed with regularized work in mind to suit GPU computation. With this work and our continuing effort to optimize our implementations of {\tt gQM3D} and {\tt gQM3D}$^+$, the computation of a quality triangulation can possibly be an integral part of interactive engineering or scientific applications. 
In addition, the approach and strategy used in this work are of independent interest to studying other variants of 3D and surface triangulation problems such as that by {\tt CGAL} and {\tt TetWild} to realize them in GPU.
%

\bibliographystyle{ACM-Reference-Format}
\bibliography{paper-bibliography}

\end{document}